\documentclass[twocolumn,showpacs,preprintnumbers,superscriptaddress,prl]{revtex4}
\usepackage{graphicx}
\usepackage{dcolumn}
\usepackage{bm}

\def\lsco{La$_{2-x}$Sr$_x$CuO$_4$}
\def\ybco{YBa$_2$Cu$_3$O$_{6+x}$}
\def\lbcoate{La$_{1.875}$Ba$_{0.125}$CuO$_{4}$}
\def\qaf{${\bf Q}_{\rm AF}$}
\def\ecross{$E_{\rm cross}$}

\begin{document}
\preprint{}

\title{Magnetic Excitations of the Diagonal Incommensurate Phase in Lightly-Doped La$_{2-x}$Sr$_{x}$CuO$_{4}$}
\author{M.~Matsuda}
\affiliation{Quantum Beam Science Directorate,
Japan Atomic Energy Agency, Tokai, Ibaraki 319-1195, Japan}
\author{M.~Fujita}
\affiliation{
Institute for Materials Research, Tohoku University, Katahira, Sendai 980-8577, 
Japan}
\author{S.~Wakimoto}
\affiliation{Quantum Beam Science Directorate,
Japan Atomic Energy Agency, Tokai, Ibaraki 319-1195, Japan}
\author{J.~A.~Fernandez-Baca}
\affiliation{
Neutron Scattering Science Division, Oak Ridge National Laboratory, Oak Ridge, 
Tennessee 37831, USA}
\author{J.~M.~Tranquada}
\affiliation{
Brookhaven National Laboratory, Upton, New York 11973, USA}
\author{K.~Yamada}
\affiliation{
Institute for Materials Research, Tohoku University, Katahira, Sendai 980-8577, 
Japan}
\date{\today}

\begin{abstract}
We present inelastic neutron scattering experiments on a single-domain crystal of lightly-doped La$_{1.96}$Sr$_{0.04}$CuO$_{4}$.  We find that the magnetic excitation spectrum in this insulating phase with a diagonal incommensurate spin modulation is remarkably similar to that in the superconducting regime, where the spin modulation is bond parallel.  In particular, we find that the dispersion slope at low energy is essentially independent of doping and temperature over a significant range.  The energy at which the excitations cross the commensurate antiferromagnetic wave vector increases roughly linearly with doping through the underdoped regime.
\end{abstract}

\maketitle

Superconductivity in the cuprates is obtained by doping holes into a parent antiferromagnetic insulator.  How the transition occurs from antiferromagnetic to superconducting ground states as a function of doping remains a controversial issue.  In the case of \lsco\ (LSCO), the commensurate antiferromagnetic order disappears at $x=0.02$, while superconductivity appears for $x\agt0.055$; the region in between is often called the ``spin-glass'' phase \cite{birg06}.  Neutron scattering studies have shown that, at low temperatures ($T<T_{\rm el}\sim20$~K), magnetic ordering is indicated by the appearance of incommensurate elastic peaks split about the antiferromagnetic wave vector, \qaf\ \cite{fuji02,waki99}.  The corresponding spin modulation occurs within the CuO$_2$ planes along a direction at 45$^\circ$  to the Cu-O bonds; in fact, it has a unique orientation with respect to the orthorhombic axes \cite{waki00}.  The magnetic incommensurability rotates to parallel to the Cu-O bond direction with the onset of superconductivity at $x\sim0.055$.   The change in character of the magnetic correlations at the onset of superconductivity is of particular interest, especially in light of recent evidence for a vortex-liquid state extending from the superconducting regime down to $x\approx0.03$ \cite{li07b}.

Here we present an inelastic neutron scattering study of the spin dynamics in LSCO with $x=0.04$, in the middle of the spin-glass regime.  Despite the rotated incommensurability, we find that the spectrum is consistent with the ``hour-glass'' dispersion observed in superconducting LSCO \cite{tran04,vign07,kofu07} (and also in \ybco\ \cite{arai99,hayd04,rezn04,stoc05,hink07}).  Even at $T\gg T_{\rm el}$, there is a strong anisotropy of the low-energy fluctuations, consistent with the nematic-like response recently reported for YBa$_2$Cu$_3$O$_{6.45}$ \cite{hink08,kive98}. Furthermore, the effective velocity of the low-energy excitations is approximately independent of energy, so that the energy, \ecross, at which the low-energy excitations cross \qaf\ scales linearly with doping, following the elastic limit of the magnetic incommensurability.  The lack of variation of the spin-fluctuation velocity is reminiscent of the doping-independence of the Fermi velocity observed in photoemission studies \cite{zhou03}; however, we can rule out a spin-density-wave instability as the source of the low-energy magnetic correlations, because the magnitude of $2k_{\rm F}$ (twice the Fermi wave vector) measured by photoemission \cite{yosh06} is inconsistent with the magnetic wave vector.  At energies above \ecross, the magnetic excitations are consistent with the effective antiferromagnetic dispersion previously observed in LSCO with $x=0.05$ \cite{goka03}.  This supports the concept that the magnetic excitations evolve directly from the antiferromagnetic insulator.

The main focus of this work involves a single crystal of La$_{1.96}$Sr$_{0.04}$CuO$_{4}$; however, we will also show some results for La$_{1.95}$Sr$_{0.05}$Cu$_{0.97}$Zn$_{0.03}$O$_{4}$ (LSCZO), which is quite similar \cite{mats06}.  Both crystals were grown by the travelling solvent floating zone method and have dimensions of  $\sim6\,\phi\times25$~mm$^{3}$.  Each crystal contains essentially a single domain of the low-temperature orthorhombic structure, with lattice parameters $a=5.34$~\AA\ and $b=5.42$~\AA\ for the pure LSCO crystal at low temperature.  The $a$ and $b$ axes are at 45$^\circ$ to the Cu-O bonds.  In this coordinate system, ${\bf Q}_{\rm AF} = (100)$, (010), in reciprocal lattice units, $(2\pi/a,2\pi/b,2\pi/c)$.  The incommensurate magnetic peaks are split uniquely along ${\bf b}^\ast$ by an amount $\epsilon$.  

The elastic magnetic scattering was characterized on the cold triple-axis spectrometer LTAS at the Japanese Research Reactor (JRR-3) of the Japan Atomic Energy Agency (JAEA).  Neutrons with an energy of 5~meV were used,  together with a horizontal collimator sequence of guide--$80'$--S--$80'$--$80'$; contamination from higher-order beams was effectively  eliminated using Be filters.  The LSCO sample yielded incommensurate peaks with $\epsilon=0.0513(7)$ rlu for $T\alt20$~K, with a Lorentzian line width of $\kappa=0.04$~\AA$^{-1}$.  For LSCZO, $\epsilon=0.0543$ rlu \cite{mats06}.

The inelastic scattering measurements on LSCO were performed on the thermal triple-axis spectrometers TAS-1 and TAS-2 at JRR-3 with a fixed final neutron energy, $E_f$, of 13.7 meV. The horizontal collimator sequences were open--$80'$--S--$80'$--$80'$ on TAS-1 and guide--$80'$--S--$80'$--open on TAS-2.  For LSCZO, the measurements were performed on the thermal triple-axis spectrometer HB-1 at the High Flux Isotope Reactor (HFIR), Oak Ridge National Laboratory. The measurement conditions were $E_f= 14.7$ meV, horizontal collimations of $48'$--$60'$--S--$80'$--$240'$. Pyrolytic graphite (PG) filters were used to suppress higher harmonics. The single crystals were oriented in the $(HK0)$ scattering plane and were mounted in a closed-cycle He gas refrigerator.

\begin{figure}
\includegraphics[width=8.5cm]{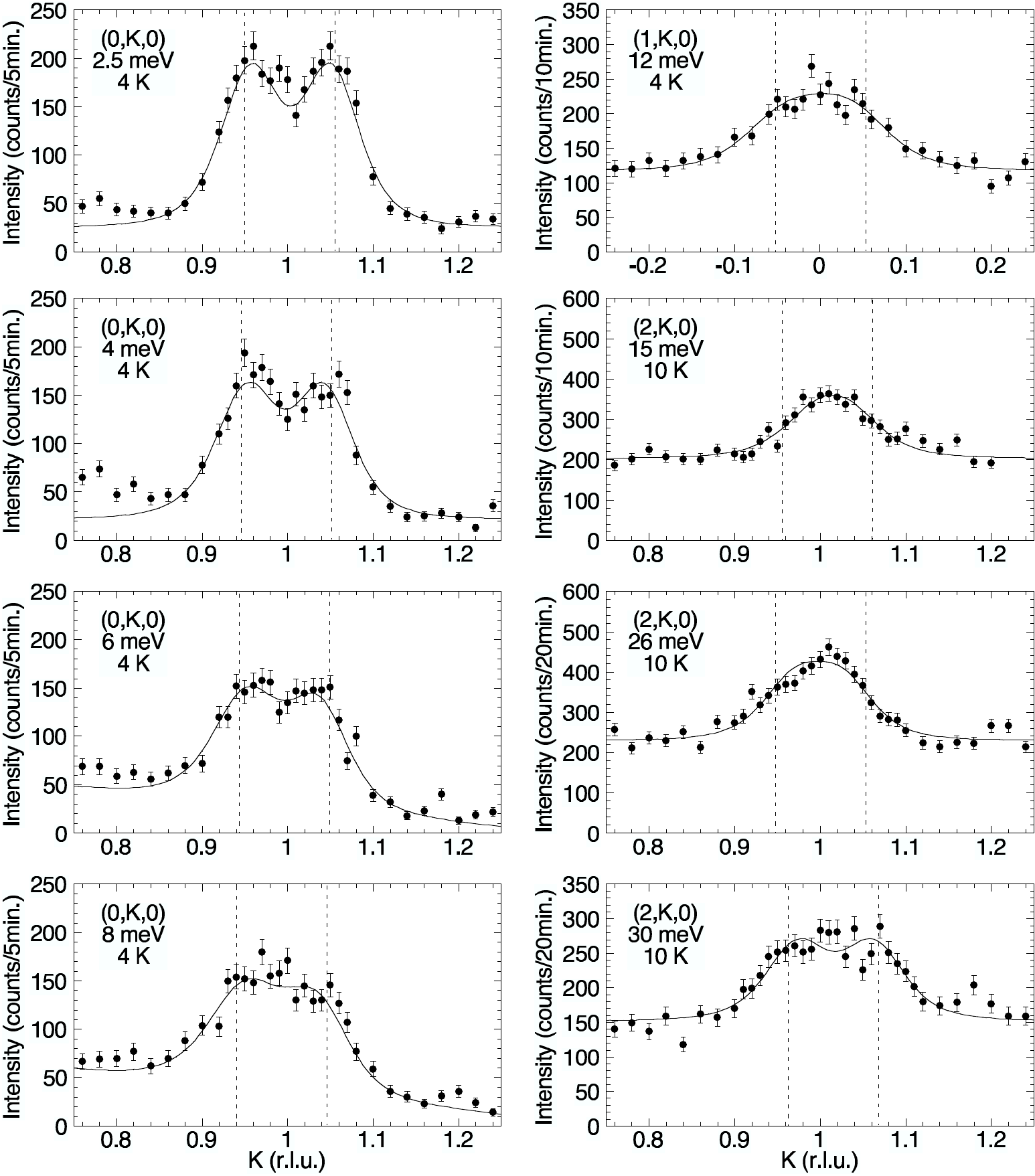}
\caption{Neutron inelastic scans along $(0,K,0)$, $(1,K,0)$, or $(2,K,0)$ in La$_{1.96}$Sr$_{0.04}$CuO$_{4}$. The solid lines are the results of fits as discussed in the text. The broken lines represent the peak positions observed at $\hbar\omega=0$ meV. The measurements were performed on TAS-2 for $\hbar\omega\le12$ meV and on TAS-1 for $\hbar\omega\ge15$ meV.} 
\label{fig1}
\end{figure}

Figure~\ref{fig1} shows the typical neutron inelastic scattering spectra between 2.5 meV and 30 meV in LSCO ($x=0.04$) measured at low temperature along the modulation direction.  At energies of 4 meV and below, the two peak structure is clear. With increasing energy, the peak separation becomes smaller and the peak width also becomes broadened. We have confirmed that the peak broadening corresponds to the resolution effect. The solid lines in Fig.~\ref{fig1} are the results of fits of a convolution of the resolution function with two 2-dimensional (2D) squared Lorentzians.  In the fitting, the linewidth ($\kappa$) of each peak, which corresponds to the effective inverse correlation length, is assumed to be isotropic in the $a$--$b$ plane and fixed at the value determined from the elastic scattering (0.04 \AA$^{-1}$). This model reproduces the observed spectra reasonably well for all measured energies ($2.5\le\hbar\omega\le30$ meV).  The analysis indicates that the peak separation (2$\epsilon$) decreases towards zero as $\hbar\omega$ is raised  to 15 meV.  Since phonon scattering contaminates the magnetic signal in the range of 15$<\hbar\omega<$25 meV, we could not clarify the magnetic spectra in that region.  The peak separation appears to grow for $\hbar\omega> 25$ meV.  The $Q$-integrated intensity decreases monotonically with increasing energy. This behavior is consistent with that observed in LSCO with $x=0.05$ \cite{goka03}; however, it is distinct from the reported behavior in LSCO with $x=0.16$ \cite{vign07} and in \lbcoate\ \cite{tran04}, where the intensity peaks at finite energy.

\begin{figure}
\includegraphics[width=6.3cm]{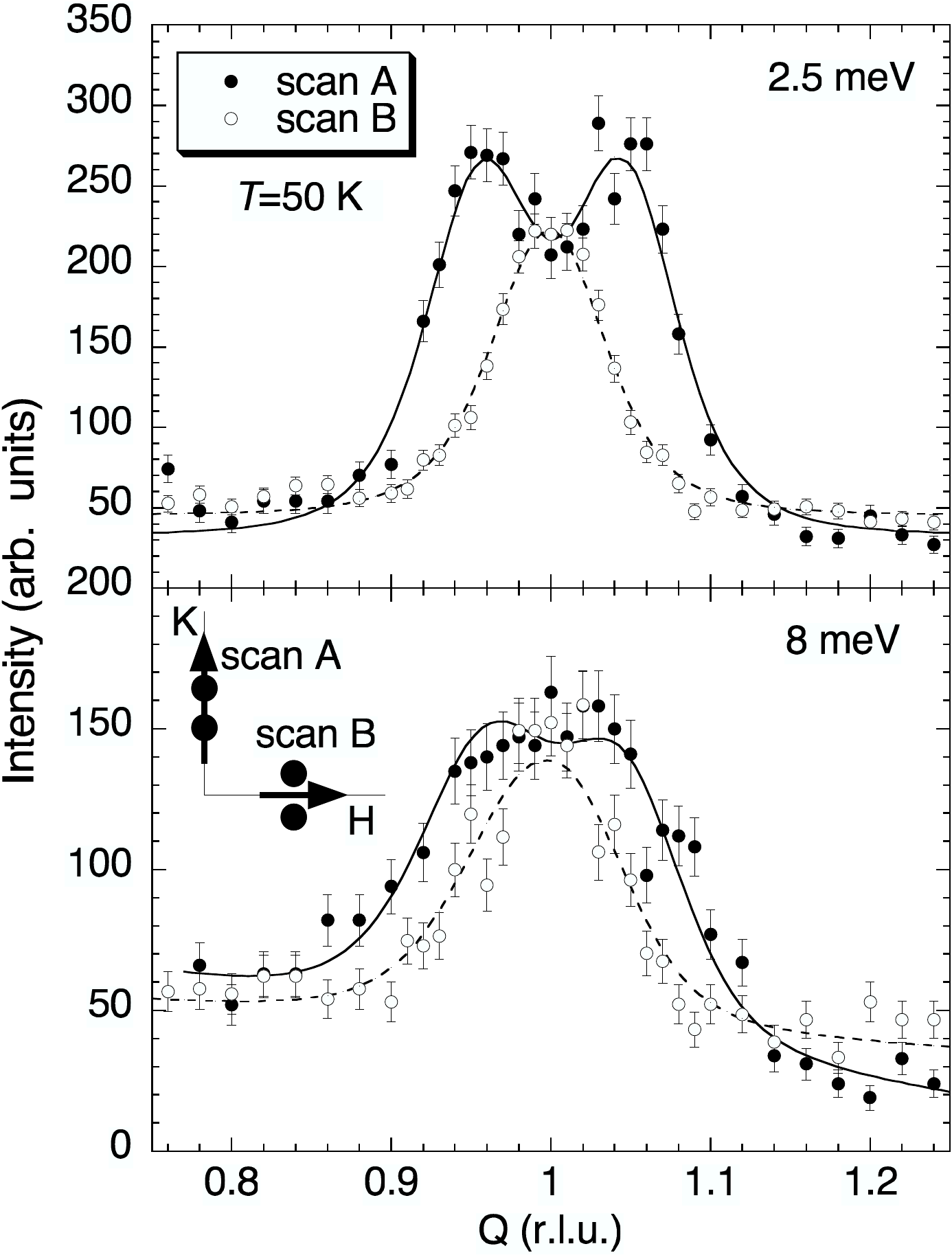}
\caption{Scans measured parallel (scan A) and perpendicular (scan B) to the incommensurate wave vector, as indicated in the inset, for energies of 2.5 and 8 meV at $T=50$~K.  The particular positions for the scans were chosen to optimize the resolution.}
\label{fig4}
\end{figure}

The anisotropy of the scattering, measured parallel and perpendicular to the incommensurate peaks, is shown in Fig.~\ref{fig4} for $T=50$~K$\null\gg T_{\rm el}$, where there is no static order.  This is to illustrate that the character of the low-energy spin correlations does not depend on whether there is a static component.  As noted in the introduction, this anisotropy looks very similar to that found recently in YBa$_2$Cu$_3$O$_{6.45}$ \cite{hink08} and consistent with expectations for a nematic electronic state \cite{kive98}.  (Of course, in both of these cases, there is already a reduction from 4-fold symmetry due to the crystal structure.)  It is also of interest to consider the 2D distribution of intensity at 25 meV and above; however, resolution effects prevented any definitive conclusions.

\begin{figure}
\includegraphics[width=6.0cm]{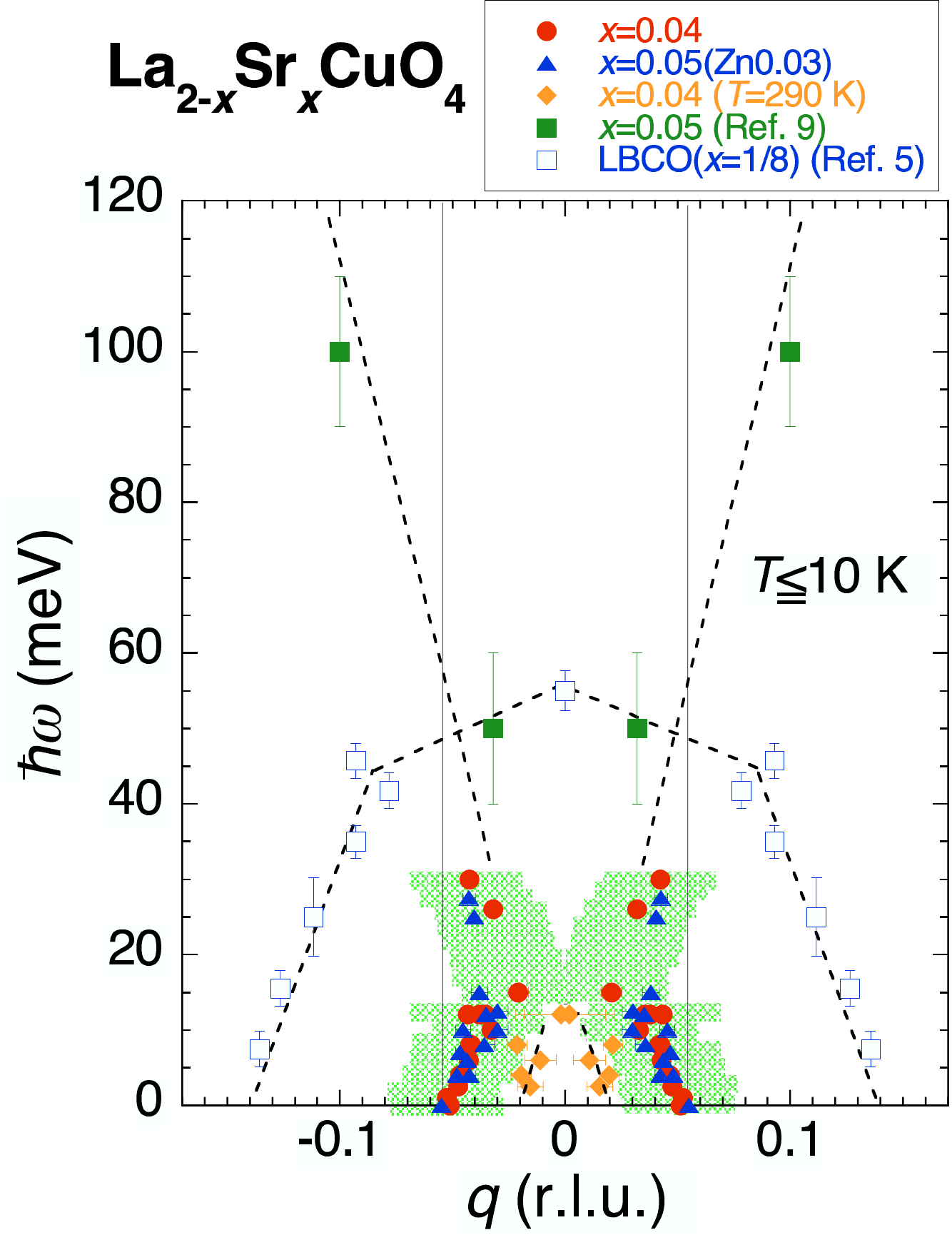}
\caption{(Color online) Magnetic dispersion relation along $q_K$ in La$_{1.96}$Sr$_{0.04}$CuO$_{4}$ (filled circles) below 10 K and at 290 K (filled diamonds). For comparison, magnetic dispersion relations in other related compounds are also shown. The filled triangles, filled squares, and open squares are data of La$_{1.95}$Sr$_{0.05}$Cu$_{0.97}$Zn$_{0.03}$O$_{4}$, La$_{1.95}$Sr$_{0.05}$CuO$_{4}$ (Ref.~\cite{goka03}), and La$_{1.875}$Ba$_{0.125}$CuO$_{4}$ (Ref.~\cite{tran04}), respectively. It  is noted that the peak positions are 45$^\circ$ rotated in La$_{1.875}$Ba$_{0.125}$CuO$_{4}$. The thick shaded bars represent the full width at half maximum of the excitation peaks in La$_{1.96}$Sr$_{0.04}$CuO$_{4}$. The interpolated data are also shown. The broken lines are visual guides.}
\label{fig2}
\end{figure}
 
Figure~\ref{fig2} summarizes the neutron inelastic results for our samples.  For $x=0.04$, the effective incommensurability decreases towards zero as the energy approaches 15 meV, while it grows for 25 meV and above.  The results for the LSCZO sample are almost identical.  At higher energies, we expect the dispersion to be similar to that measured in \lsco\ with $x=0.05$ \cite{goka03}, and these points are included in the figure.  The lower energy results for La$_{1.875}$Ba$_{0.125}$CuO$_{4}$~\cite{tran04} are also shown for comparison (plotted as a function of $|{\bf q}|$, and ignoring the difference in orientation of {\bf q}).   Overall, the magnetic dispersion has an hour-glass shape similar to that observed in the parallel incommensurate phase.  

\begin{figure}
\includegraphics[width=6.3cm]{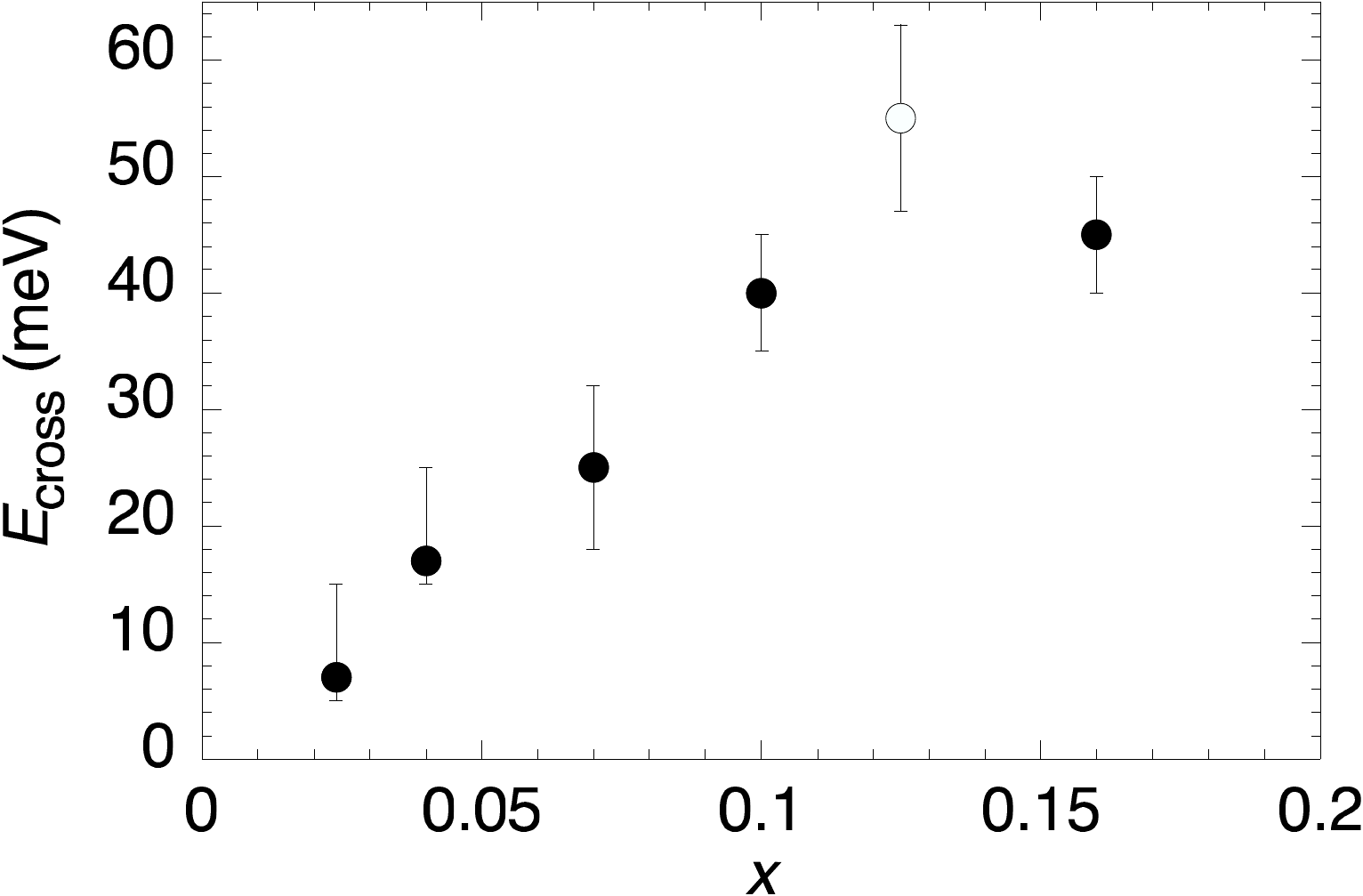}
\caption{Plot of \ecross\ vs.\ $x$ in LSCO (filled circles) and La$_{2-x}$Ba$_x$CuO$_4$ (open circles), including results from Refs.~\cite{mats00,hira01,kofu07,tran04,vign07}.}
\label{figEc}
\end{figure}

An intriguing feature is that the slope of the effective low-energy dispersion (which corresponds to a velocity) does not appear to depend strongly on doping.  If the velocity is independent of doping, while the incommensurability is linear in $x$ for $x\alt\frac18$, then one would expect \ecross\ to be proportional to $x$ \cite{bati01}.  In Fig.~\ref{figEc}, we plot values for \ecross\, with conservative error bars, extracted from a series of studies.  Indeed, we observe that $E_{\rm cross} \sim x$ for $x\alt\frac18$.  This clear trend suggests that there must be some degree of continuity of the magnetic correlations across the insulator-to-superconductor transition.

\begin{figure}
\includegraphics[width=6.3cm]{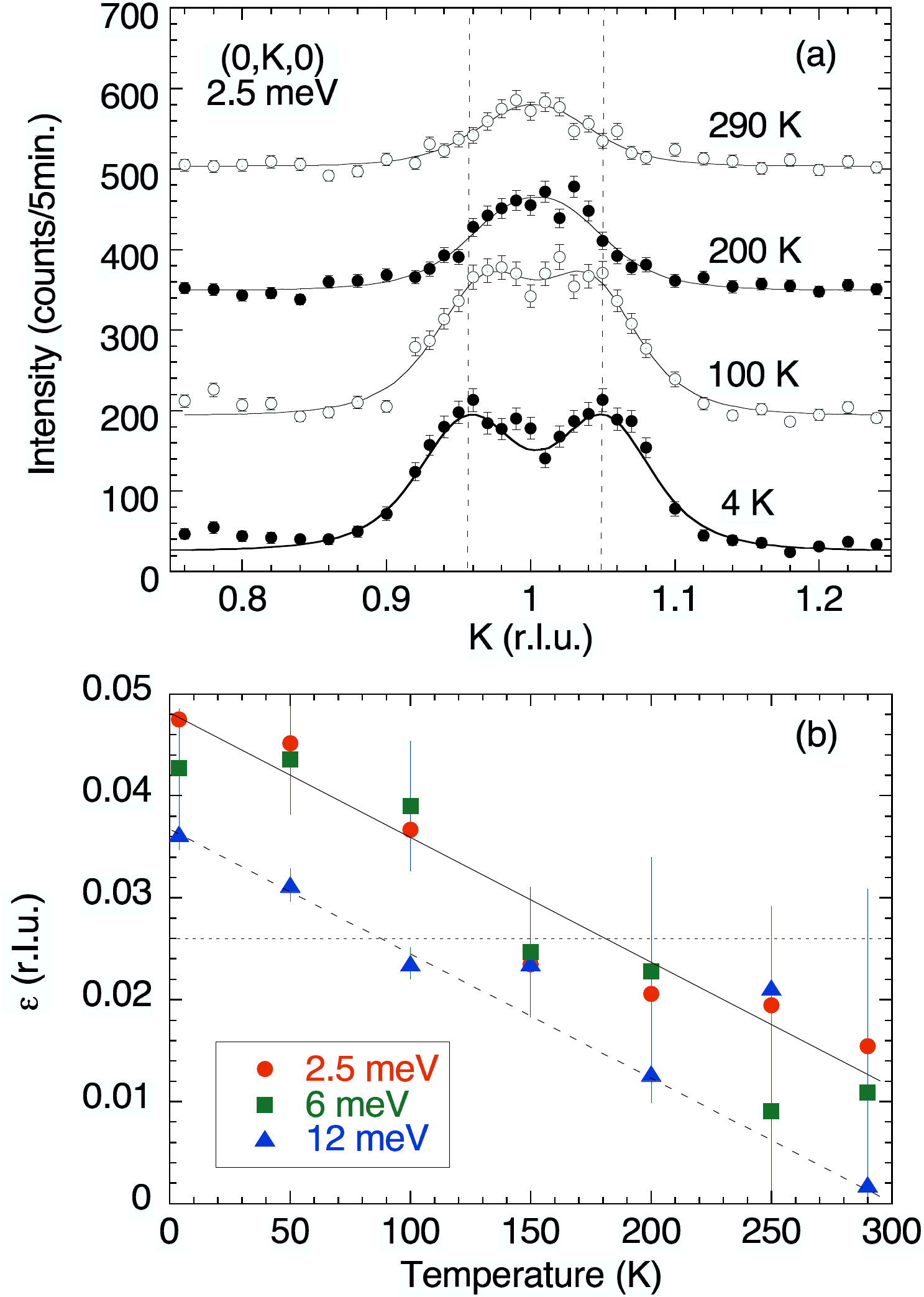}
\caption{(Color online) (a) Temperature dependence of the magnetic excitation 
spectra at 2.5 meV. Successive scans have been displaced vertically by 150 counts 
for clarify. The solid lines are the results of fits of a convolution of the 
resolution function with two 2D squared Lorentzians. The broken lines represent 
the peak positions at 4 K. (b) Temperature dependence of the incommensurability 
($\epsilon$) at 2.5, 6, and 12 meV in La$_{1.96}$Sr$_{0.04}$CuO$_{4}$. The solid and broken lines are visual guides for the data. The dotted horizontal line shows the peak width (HWHM). Two peaks can hardly be resolved when $\epsilon$ becomes much smaller than this value.}
\label{fig3}
\end{figure}

We also investigated the temperature dependence of the magnetic excitation spectra.  Figure~\ref{fig3}(a)  shows scans at  $\hbar\omega=2.5$ meV.   The lines through the data are similar fits to those in Fig.~\ref{fig1}, now with $\kappa$ taken to be temperature independent.  (Even if $\kappa$ were temperature dependent, we estimate an upper limit at 290~K of $\sim0.05$~\AA$^{-1}$, an increase of only 25\%.)  We conclude that $\epsilon$ gradually decreases as the temperature rises, with the intensity also decreasing.  The temperature dependence of $\epsilon$ evaluated at 2.5, 6, and 12 meV is plotted in Fig.~\ref{fig3}(b).  In each case, the decrease is roughly linear in temperature.  A consequence of this is that the slope of the dispersion does not change; note the effective dispersion at 290 K indicated in Fig.~\ref{fig2}.

Just as the incommensurability evolves with temperature, so does the electronic conductivity.  In particular, the conductivity is poor at room temperature where the magnetic correlations are almost commensurate. The low-frequency optical conductivity grows with cooling \cite{dumm03}, as the spin incommensurability develops.  Given the continuous evolution of \ecross\ with doping, providing a connection with La$_{1.875}$Ba$_{0.125}$CuO$_4$ where charge and spin stripe order is known to occur \cite{tran04}, we believe that charge stripes and moment modulation are likely to be an important part of the incommensurate reponse in LSCO with $x=0.04$.  We are aware that there has been considerable work on spiral-order models for the spin-glass phase \cite{lusc07,hass04,berc04,juri06,brug07}.  While we cannot rule out such models on the basis of the observed magnetic spectrum, explaining Fig.~\ref{figEc} may be a challenge for such an approach.

Another possibility is spin-density-wave order due to Fermi-surface nesting \cite{frie89}.  Angle-resolved photoemission measurements on LSCO \cite{yosh06} indicate that $2k_{\rm F}$ in the nodal direction would yield a diagonal magnetic modulation with $\epsilon=0.15$~\AA$^{-1}$ for $x=0.04$.  This happens to be 2.5 times greater than the experimental value of 0.06~\AA$^{-1}$.  Hence, a simple nesting picture appears to be ruled out.

In terms of a stripe picture, the dominant magnetic interaction would be superexchange within locally antiferromagnetic domains.  There is still a challenge to understand why the dispersion of the low-energy excitations is not signficantly affected by the rotation in stripe orientation.  One possibility suggested by Granath \cite{gran04} is that a diagonal stripe might consist of a staircase pattern of bond-parallel stripes.  He found that such a pattern seemed necessary in order to obtain consistency with the photoemission experiments \cite{yosh06}.   We note that the magnetic spectrum is distinct from that observed in the insulating diagonal stripe phase of La$_{2-x}$Sr$_x$NiO$_4$ \cite{bour03,woo05}.  Also, $E_{\rm cross} \sim x$ is a stronger variation than predicted by the stripe-phase calculations of Seibold and Lorenzana \cite{seib06}.  Thus, while our results may be phenomenologically consistent with stripe correlations, improvements in theoretical models are needed.

We would like to thank T. Tohyama and Y. Koike for stimulating discussions. This study was supported in part by the U.S.-Japan Cooperative Program on Neutron Scattering and by a Grant-in-Aid for Scientific Research from the MEXT of Japan. Work at Oak Ridge National Laboratory and Brookhaven National Laboratory was supported by the U.S. Department of Energy's Office of Science under Contract Nos. DE-AC05-00OR22725A and DE-AC02-98CH10886, respectively.


\end{document}